\documentstyle[aps,prl,epsf,multicol]{revtex}
\epsfclipon
\begin{document} 
\draft
\title{Critical Coarsening without Surface Tension: the Voter Universality Class}
\author{Ivan Dornic,$^1$ Hugues Chat\'e,$^2$
 J\'er\^ome Chave,$^3$  and Haye Hinrichsen $^4$}

\address{
$^1$Max-Planck-Institut f\"ur Physik komplexer Systeme,
N\"othnitzer Stra\ss e 38, 01187 Dresden, Germany\\
$^2$CEA -- Service de Physique de l'\'Etat Condens\'e, 
Centre d'\'Etudes de Saclay, 91191 Gif-sur-Yvette, France\\
$^3$Department of Ecology and Evolutionary Biology, 
Princeton University, Princeton NJ 08544-1003, USA\\
$^4$Theoretische Physik, Fachbereich 8, 
Universit\"at GH Wuppertal, Gau\ss stra\ss e 20, 42097 Wuppertal, Germany}
\date{14 January 2001}   
\maketitle

\begin{abstract}
We show that the two-dimensional voter model, usually considered to only be a
marginal coarsening system, represents a broad 
class of models for which phase-ordering takes place without surface tension.
We argue that voter-like growth is generically observed at 
order-disorder nonequilibrium transitions solely driven by interfacial
noise between dynamically symmetric absorbing states.
\end{abstract}


\begin{multicols}{2}

\begin{figure}[htbp]
\narrowtext
\centerline{
\epsfxsize=8.cm
\epsffile{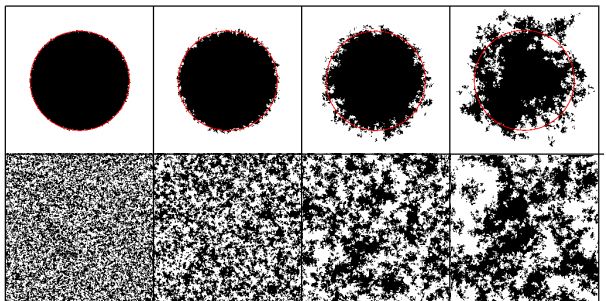}
}
\caption{Domain growth in the  usual VM  (system size $256^2$).
Top: snapshots at $t=4, 16, 64, 256$ during the evolution of a 
bubble of initial radius $r_0=180$ (thin circle). 
Bottom: same from symmetric r.i.c.}
\label{f1}
\end{figure}

Coarsening phenomena occur in a large variety of situations 
in and out of physics, ranging from the demixion of alloys~\cite{Bra94} 
to population dynamics~\cite{Mol99}.
At a fundamental level,  phase-ordering
challenges our capacity to deal with nonequilibrium systems and
our understanding of  the mechanisms determining different
universality classes.
In many cases, phase competition is driven by surface tension,
leading to `curvature-driven' growth. Coarsening patterns are
then characterized by a single length scale $L(t) \sim t^{1/z}$,
where the exponent $z$ only depends on general symmetry and 
conservation properties of the system~\cite{Bra94}. For instance, $z=2$ for
the common case of a non-conserved scalar order parameter (NCOP), 
a large class including the Ising model.
In this context, the two-dimensional voter model (VM)~\cite{Lig85}, 
a caricatural process in which sites 
on a square lattice adopt the opinion of a randomly-chosen neighbor,
stands out as an exception.
Its coarsening process, which gives rise to
patterns with clusters of all sizes between 1 and $\sqrt{t}$
(Fig.~\ref{f1})~\cite{Cox86,Sch88}, is characterized by a slow,
logarithmic decay of the density of interfaces
$\rho \sim 1/\ln t$ (as opposed to the algebraic decay
$\rho \sim 1/L(t) \sim t^{-1/z}$ of curvature-driven growth).
The marginality of the VM is usually attributed to the exceptional
 character of its analytic properties~\cite{Cox86,Sch88,Kra92}. 

In this Letter, we show that, in fact, large classes of models exhibit 
the same type of domain growth as the simple VM, without
being endowed with any of its peculiar symmetry and integrability
properties. We argue that voter-like coarsening
is best defined by the absence of surface tension and that it is generically
observed at the transitions between disordered and fully-ordered
phases in the absence of bulk fluctuations, when these 
nonequilibrium transitions are driven by interfacial noise only. 
Finally, we discuss the universality of the scaling properties associated 
with voter-like critical points.

We first review the properties of the usual two-state VM,
 emphasizing those of importance for defining
generalized models. A `voter' (or spin) residing on site {\bf x} of a
hypercubic lattice can have two 
different opinions $s_{\bf x}=\pm 1$. In any space dimension $d$,
an elementary move consists in randomly
choosing one site and assigning to it the opinion of one of its randomly
chosen nearest neighbors (n.n.). This  ensures that the two homogeneous
configurations (where all spins are either $+$ or $-$)
are absorbing states, and that
the model is $Z_2$-symmetric under global inversion 
($s_{\bf x} \to - s_{\bf x}$).
Recasting the dynamic rule  in terms of n.n. pair updating,
a pair of opposite spins  $+-$ is randomly selected, and 
  evolves to a $++$ or $--$ pair with
equal ($1/2$) probabilities.  
The rates of creation of $+$ and $-$ 
spins  being equal, any initial value of the global magnetization $m$
is conserved in the limit of large  system sizes.

Another prominent feature of the VM
is a `duality'~\cite{Lig85} with a system of coalescing  random walks:
going backward in time, the successive ancestors of a given spin 
follow the trail of  a simple random walk (RW); comparing the values
 of several spins shows that their associated RWs necessarily 
merge upon encounter. This correspondence allows to solve
many aspects of the kinetics of the VM,  because
it implies that the correlation  functions between
 an arbitrary number of spins form a closed hierarchy of 
diffusion equations~\cite{Sch88,Kra92}. In particular,
 the calculation of the density of interfaces $\rho_m(t)$
(i.e. the fraction of $+-$ n.n. pairs)
starting from  random initial conditions (r.i.c.) 
of magnetization $m$, is ultimately given by
the  probability that a  RW  initially at unit distance 
from the origin, has not yet reached it at time $t$.
Therefore, owing to the recurrence properties of RWs, 
the VM shows coarsening for $d \le 2$
(i.e. $\rho_{m}(t)\to 0$ when $t \to \infty$).
For the `marginal' case $d=2$
---on which we mainly focus henceforth---,
one finds the slow logarithmic decay~\cite{Sch88,Kra92}:
\begin{equation}
\rho_m(t) = (1-m^2)\left[\frac{2 \pi D}{\ln t} +
{\cal O}\left(\frac{1}{\ln^2{t}}\right)\right] \;,
\label{eq_rho}
\end{equation}
 $D$  being the diffusion constant of the underlying RW 
($D=1/4$ for the standard case of n.n., square lattice walks, 
when each spin is updated on average once per unit of time).
This peculiar behavior, which contrasts 
 with the algebraic decay encountered in NCOP growth, thus seems 
tantamount to
the fact that $d=2$ is the critical dimension for the recurrence of RWs.

Beyond these analytic properties,
the nature of the coarsening displayed by the VM in two dimensions 
is highlighted
by studying the evolution of a `bubble' of one phase embedded into another.
For curvature-driven NCOP growth, the volume of a bubble
decreases linearly with time under the effect of surface tension. 
In the VM, however,   large bubbles do not shrink but slowly
disintegrate as their  boundary roughens diffusively to reach a typical
width of the order of  their initial radius $r_0$ (Fig.~\ref{f1}, top).
As long as $\sqrt{t} \ll r_0$, 
conservation of the magnetization is still effective, 
and implies  that radially-averaged magnetization profiles $m(r,t)$
have a stationary middle point.
This indicates that curvature plays no role, a fact further confirmed 
by the observation that the $m(r,t)$ curves are the same as the
transverse profiles of an infinite, straight interface,
whose derivatives can be shown to be given 
by a simple Gaussian of variance $2Dt$.

We interpret the above behavior  as characteristic of the absence 
of surface tension, a `physical' property which, together 
with the fact that coarsening does occur (i.e., $\lim_{t\to\infty} \rho =0$), 
we conjecture to be  constitutive of $d=2$ voter-like domain growth 
(as defined by the logarithmic decay of $\rho_m(t)$). 
To investigate which of the properties 
outlined above (existence of two absorbing states, 
$Z_2$-symmetry, duality with coalescing RWs, $m$-conservation) is necessary to
produce voter-like growth, we now embed the simple VM studied above
into large families of rules.

Consider first that a $+-$ pair is now only updated with probability
$p$ (${\rm Pr}(+-\to +-)=1-p$), and that ${\rm Pr}(+-\to ++)=pq$ and 
 ${\rm Pr}(+-\to --)=p(1-q)$, so that for $q\ne\frac{1}{2}$ the
$m$-conservation is broken. Suppose next that $p$ and $q$
depend on the local configuration around the $+-$ pair,
for instance via $n^+$ and $n^-$, the numbers of  
$+$ (resp. $-$) n.n. of the
$-$ (resp. $+$) site in the central pair.
(Note that as soon as the transition probabilities 
do vary with the local
configuration, the duality with coalescing RWs breaks down.)
Conservation of $m$ (Pr($+-\to++$) = Pr($+-\to --$))
is then simply expressed by the conditions $q_{n^+,n^-} = \frac{1}{2}$, 
to be obeyed $\forall \, n^+ \!,n^- \in \{1,2,3,4\}$, whereas
$p_{n^-,n^+} = p_{n^+,n^-}$ and $q_{n^-,n^+} = 1 - q_{n^+,n^-}$ stand
for $Z_2$-symmetry.

With the exception of the usual VM ($p=1, q=1/2$),
the $m$-conserving and $Z_2$-symmetric rules, defined by
$q_{n^+,n^-} = 1/2$ and $p_{n^-,n^+} = p_{n^+,n^-}$,
do {\it not} benefit
from the duality property used above to derive, e.g., 
Eq.~(\ref{eq_rho}). Nevertheless,
we have found (by coarsening,
bubble and line experiments \cite{Ustocome}) that they 
behave like the simple VM, but with a diffusion constant
$D\ne \frac{1}{4}$.
Thus, integrability is not a necessary condition for voter-like growth.

\begin{figure}[htbp]
\narrowtext
\centerline{
\epsfxsize=8.cm
\epsffile{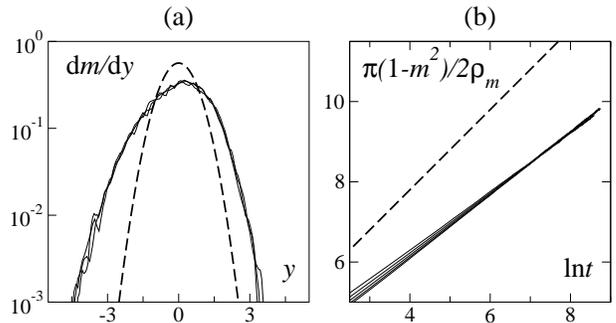}
}
\caption{Domain growth in the KKE model and the n.n. ($D=\frac{1}{4}$) VM.
(a):  derivative of rescaled magnetization profiles at  times $t=16,64,256$ 
(data from a bubble experiment;
initial conditions: $m(r,0)={\rm sign}(r-r_0)$ with $r_0=4096$).
Dashed line:  exact result for the VM
${\rm d}m/{\rm d}y=(2/\sqrt{\pi}) e^{-y^{2}}$ with $y=(r-r_0)/\sqrt{t}$;
(b): scaling of $\rho_m(t)$ (as suggested by Eq.~(\protect{\ref{eq_rho}}))
in coarsening experiments 
for  $m=0,\pm0.2,\pm0.4$ in systems of size $16384^2$; dashed line: exact result
for the VM.}
\label{f2}
\end{figure}

The subset of $m$-conserving rules without $Z_2$-symmetry is 
of particular interest: no surface tension is expected, but 
one may wonder about the effects of the asymmetry on growth properties.
Below, we use the case $p_{n^+,n^-}=1/n^+$, which is nothing but
the interface growth model introduced by 
Kaya, Kabak{\c{c}}{\i}o{\u{g}}lu, and Erzan (KKE)
at its `delocalization' transition~\cite{Kke00}. 
The asymmetry of the rule manifests itself at
the microscopic level (by looking at local 3-spins configurations
\cite{Ustocome}),
but also at the macroscopic level, by the evolution of the
magnetization profiles in `line' or `bubble' experiments. As for the usual VM,
the profiles obtained in both cases are identical, their width
scale as $2 D_{\rm KKE}\sqrt{t}$ with 
$D_{\rm KKE} = 0.36(2)$ \cite{Note1}, but they
are asymmetric (Fig.~\ref{f2}a).
 Voter-like behavior is further confirmed by the study of phase
ordering following r.i.c. of magnetization $m$. The density $\rho_m(t)$  
displays the signature scaling  spelled out by Eq.~(\ref{eq_rho}) 
(Fig.~\ref{f2}b), and the ensuing value  $D = 0.34(1)$ is
in fair agreement with the value $D_{\rm KKE}$ determined above, 
  which suggests
 that the KKE rule asymptotically behaves like the usual VM with an
{\it effective} diffusion constant.

All the other $m$-conserving, asymmetric rules that we have studied 
reveal  a similar voter-like behavior. 
We conjecture that they almost all do so, as long as   the
  $p_{n^+,n^-}$ are all strictly positive  
 (to avoid  possible `blocking' in certain configurations).
 In other words,
 $m$-conservation appears as a sufficient condition for voter-type growth, 
even in the absence of $Z_2$-symmetry or integrability. 
We now show that $m$-conservation is not
 a necessary condition since
some non-conserving $Z_2$-symmetric rules 
with two absorbing states exhibit voter-like features.

If pair-update rules are convenient to control $m$-conservation,
single-site
update models suffice to study rules which need not possess this property.
We thus consider a family of `kinetic Ising models', in which a spin 
is flipped with a probability
 $r_{s,h}$ depending on its
value $s=\pm 1$ and on its local field $h\in \{-4,-2,0,2,4\}$, 
i.e. the sum of the values of its four n.n..
The $Z_2$-symmetric rules of this family, 
obeying $r_{-s,-h}=r_{s,h}$, can be parameterized
by five transition probabilities $r_h  \equiv r_{1,h}$. 
Such rules are usually not integrable. Again, the VM,
 given by $r_h=\frac{1}{2}-\frac{h}{8}$, is an exception.
Its marginal situation in this context has already 
been noted~\cite{DeO93,Dro99} 
in the  particular case where 
the transition probabilities   only depend on $h$, 
a condition expressed here by $r_{-s,h}=1-r_{s,h}$. 
One is then left with only two free parameters,
$p_{\rm b} \equiv r_4, \ p_{\rm i} \equiv r_2$, 
which can be respectively interpreted~\cite{Dro99} as a measure of   
the strength of  `bulk' and `interfacial' noise. 
In the $(p_{\rm i},p_{\rm b})$ plane, the VM 
($p_{\rm i}= \frac{1}{4}$, $p_{\rm b} =0$) turns out to be the endpoint
of a line of Ising-like (second-order) phase transitions~\cite{DeO93,Dro99}, 
at the border between the ordered and disordered  phases in the absence of 
bulk noise ($p_{\rm b}=0$).

\begin{figure}[htbp]
\narrowtext
\centerline{
\epsfxsize=8.cm 
\epsffile{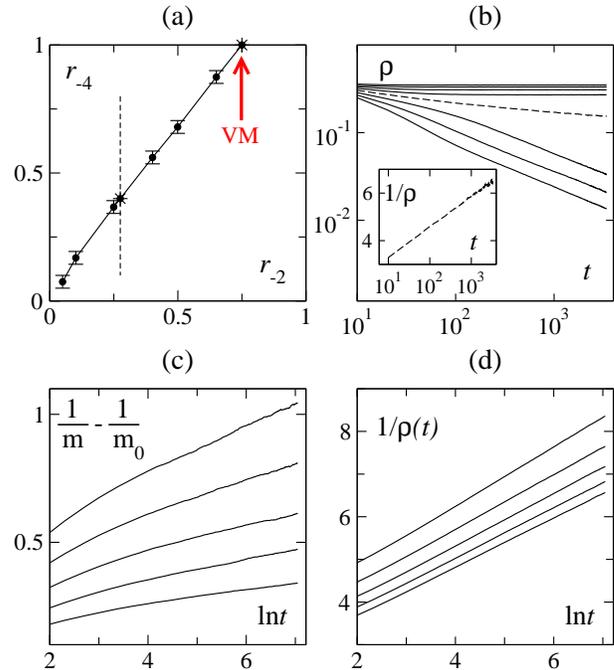}
}
\caption{Voter-like rules not conserving $m$: 
(a): phase diagram in the plane $(r_{-2},r_{-4})$ for  
$r_{0}=\frac{1}{2}, r_2=r_{-4}/4$.
(b): $\rho(t)$ from  $m_0=0$ r.i.c.
increasing $r_{-4}$ (from top to bottom) with $r_{-2}=0.275$ 
(dashed line in (a)).
Inset: interface density voter-like decay at the critical point
(system size $2048^2$).
(c,d): $m(t)$ (c) and $\rho(t)$ (d)
from r.i.c. with various $m_0$ at the critical point described in (b) 
(system size $16384^2$). 
}
\label{f3}
\end{figure}

We claim that this situation is generic: the 4-parameter 
space of the general kinetic Ising model above without bulk noise 
(space defined by $r_4=0$) is divided into a disordered 
and an ordered region separated by a codimension-1 {\it critical} manifold of 
voter-like rules~\cite{Meh99}. In Fig.~\ref{f3}, we show a section of this
 manifold in a plane which contains the VM.
Its location was determined by studying the evolution of $\rho$
following r.i.c. with zero magnetization at various parameter values. 
Crossing the  critical manifold, 
the characteristic logarithmic decay $\rho \propto 1/\ln t$ 
separates the region where $\rho \propto 1/\sqrt{t}$ (curvature-driven growth
typical of the ordered phase) from the disordered region where $\rho$ 
saturates to finite values (Fig.~\ref{f3}b).  
A peculiar feature of this voter-like critical manifold
is that, using the language of the renormalization group (RG),
$m=0$ is a `weakly attracting' fixed point of the
 dynamics: starting from any $m_0\ne 0, \pm 1$ ($m_0=m(0)$),
 one  observes that 
$m(t) \propto 1/\ln t$ (Fig.~\ref{f3}c). Yet, the decay of the interface 
density behaves like in the usual VM (Fig.~\ref{f3}d), even for 
$m_0\ne 0$, because the slow evolution of $m(t)$
introduces only  ${\cal O}(1/\ln^2 t)$ corrections 
to Eq.~(\ref{eq_rho})
and therefore does not alter the 
value of the effective diffusion constant.
This also explains why our  bubble and line experiments 
---which  strictly speaking  probe only
a zero-measure set  of random initial conditions---
performed for rules on the voter-like manifold, do
give the same results as for the usual VM \cite{Ustocome},
in spite of the absence of $m$-conservation. 

Similar investigations have been undertaken in various planes of the 
4-parameter space, always yielding the same results. This suggests 
to conjecture 
that all critical $Z_2$-symmetric rules without bulk noise
form a codimension-1 `voter-like'  manifold separating order from disorder, 
characterized by the logarithmic decay of both $\rho$ and $m$.

Does the above picture extend to non-conserving,
asymmetric rules without bulk noise?  In this case, even though 
two absorbing states exist, their dynamical roles are not symmetric,
and  such order-disorder transitions are known~\cite{J_G81,Hay00}
to belong generically to  the universality class of directed percolation (DP).
We have indeed observed~\cite{Ustocome} that for such rules
both $m$ and $\rho$ 
scale at criticality  with the {\it same} exponent 
 $\beta^{\rm DP}/\nu^{\rm DP}_\parallel = 0.450(1)$~\cite{Hay00}.

\begin{figure}[htbp]
\narrowtext
\centerline{
\epsfxsize=8.cm
\epsffile{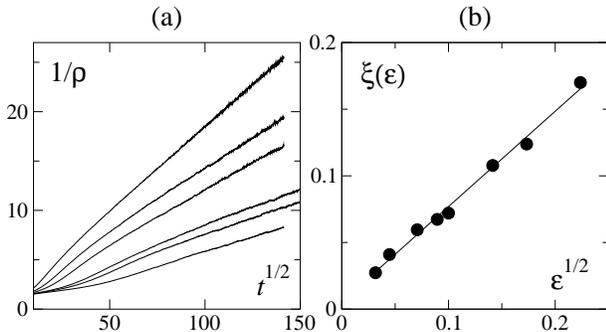}}
\caption{Phase-ordering from $m=0$ r.i.c. approaching
 the VM from the ordered side  
(system size $8192^2$).
(a) $\rho$ vs $\sqrt{t}$ for various $\varepsilon$ between 
$10^{-3}$ and $5.10^{-2}$ 
(from bottom to top);
(b) extracted correlation length $\xi$ vs $\sqrt{\varepsilon}$.}
\label{f4}
\end{figure}

Having specified the required conditions to observe voter-like
coarsening, we now turn to the
investigation of the scaling laws associated to voter-like points and
to an assessment of their universality. 
To our knowledge, there has been only one attempt 
to determining 
the scaling exponents governing the approach to the usual VM.
Within the reduced $(p_{\rm i},p_{\rm b})$ parameter plane defined above,
the authors of~\cite{DeO93}  noted first that increasing $p_{\rm i}$ 
along the zero-bulk-noise line $p_{\rm b}=0$, up to the VM
point ($p_{\rm i}=\frac{1}{4}$), the (static) 
magnetization $m_{\rm s}$ jumps from 0 to $\pm 1$.
Thus, the order parameter exponent $\beta=0$, and, for such a first-order
$d=2$ nonequilibrium critical point,  
the susceptibility and the correlation length exponents should
satisfy $\gamma=2\nu$.
A finite-size scaling (FSS) analysis of the fluctuations of $m_s$ as 
$\varepsilon \equiv p_{\rm i}-\frac{1}{4} \to 0^{-}$ confirmed the latter
relation, and gave 
 $\gamma \approx 1.25$. 
However, repeating these simulations with much better statistics, 
we evidenced \cite{Ustocome} a  systematic
decrease of the local exponent $\gamma$ as $\varepsilon \to 0^-$. 
On general grounds, corrections to scaling in this problem
are expected to be logarithmic \cite{Note2} and, thus, the above  $\gamma$
value must be taken cautiously.
In fact, a more reliable approach of the 
VM is from the ordered side ($\varepsilon \to 0^+$),
along the zero-bulk noise line.
The curvature-driven-growth regimes
$\rho \sim \xi(\varepsilon)/L(t) \sim \varepsilon^{-\nu}/t^{1/2}$, 
which eventually settle then in large enough systems,
are less prone to the logarithmic corrections mentioned above.
The scaling of the correlation length $\xi(\varepsilon)$ 
yields $\nu=0.45(7)$, compatible with
the simple `diffusive' value $\nu=\frac{1}{2}$ (Fig.~\ref{f4}). 

Compounding the above results,
a  conservative interpretation of the data is that
$\beta=0$, $\gamma=1$, and $\nu=\frac{1}{2}$, with appropriate
logarithmic corrections to scaling. This is further confirmed
by similar, preliminary, results obtained when
approaching any of the voter-like models studied above~\cite{Ustocome}.
We finally note  that these exponents  would obey all 
the  standard scaling relations valid at a  critical 
point, such as  
$\beta=[d-\lambda(H)]/\lambda(T_i)$, 
$\gamma=[2 \lambda(H)-d]/\lambda(T_i)$, or $\nu=1/\lambda(T_i)$,
with (in $d=2$)  $\lambda(T_i)=\lambda(H)=2$. 
We tentatively associate
the former  RG-eigenvalue with the relevance of interfacial noise,
 and the latter with
the presence of a `dynamically self-induced' magnetic field
breaking the $Z_2$-symmetry (as happens in KKE-like models).
The behavior of the classes of  
voter-like models we have 
defined is consistent with such a set of exponents, which would thus
characterize (in $d=2$) the genuine voter critical point.
 For instance, converting 
the static critical behavior $m \sim \varepsilon^\beta$
to a dynamical one via 
$\xi \sim \varepsilon^{-\nu} \sim L(t) \sim t^{1/z}$ gives
$m \sim t^{-\beta/\nu z}$.
Now, if $\beta=0$ strictly, this implies $m(t)={\rm Cst}$, while 
if $\beta=0^+$ is interpreted, as is customary, including logarithmic 
corrections, then $m(t) \sim 1/\ln{t}$. 
It is also nice to note that, despite the presence of the {\it a priori}
 relevant eigenvalue $\lambda(H) =2$, $m$-conserving
rules without $Z_2$-symmetry may still display voter-like growth,
because the critical exponent $\delta$ (which
describes how the order parameter  behaves under a magnetic field:
$m \sim H^{1/\delta}$), would be formally given by
$1/\delta = -1+d/\lambda(H)=0$.
 
Naturally, the RG picture sketched above needs to be
substantiated by the study of an appropriate field theory, an endeavor 
left for future work. 
In particular, one would like to have a better understanding of
 the RG flow around the voter critical point as the space dimension varies,
and of the special role played by $d=2$, 
which appears as the {\it upper critical dimension} where voter-like 
phase-ordering can be induced by the sole presence
of interfacial noise~\cite{Dic96,Ustocome}.
At any rate, we believe 
the results presented here reveal that the voter model, which
was, up to now, perceived as a marginal system, embodies 
in fact a broad class of models for which coarsening occurs in
the absence of 
surface tension. Our findings also suggest that
the critical behavior associated to this class of systems is perhaps
best characterized as an order-disorder transition  driven
by the interfacial noise between two absorbing states possessing
equivalent `dynamical roles', this symmetry being enforced either
by $Z_2$-symmetry of the local rules, or by the global conservation
of magnetization.

\end{multicols}
\end{document}